\documentclass{article}

\usepackage{PRIMEarxiv}

\usepackage[utf8]{inputenc} 
\usepackage[T1]{fontenc}    
\usepackage{hyperref}       
\usepackage{url}            
\usepackage{booktabs}       
\usepackage{amsfonts}       
\usepackage{nicefrac}       
\usepackage{microtype}      
\usepackage{lipsum}
\usepackage{fancyhdr}       
\usepackage{graphicx}       
\graphicspath{{Figures/}}     
\usepackage{natbib}
\usepackage{subfigure}
\usepackage{amsmath}
\usepackage{amssymb}

\title{Propagating the prior from shallow to deep with a pre-trained velocity-model Generative Transformer network
}

\author{
  Randy Harsuko, Shijun Cheng, Tariq Alkhalifah \\
  Earth Science and Engineering \\
  King Abdullah University of Science and Technology \\
  Thuwal, Saudi Arabia\\
  \texttt{\{mochammad.randycaesario, shijun.cheng, tariq.alkhalifah\}@kaust.edu.sa} \\
}

\begin{document}
\maketitle

\begin{abstract}
Building subsurface velocity models is essential to our goals in utilizing seismic data for Earth discovery and exploration, as well as monitoring. With the dawn of machine learning, these velocity models (or, more precisely, their distribution) can be stored accurately and efficiently in a generative model. These stored velocity model distributions can be utilized to regularize or quantify uncertainties in inverse problems, like full waveform inversion. However, most generators, like normalizing flows or diffusion models, treat the image (velocity model) uniformly, disregarding spatial dependencies and resolution changes with respect to the observation locations. To address this weakness, we introduce VelocityGPT, a novel implementation that utilizes Transformer decoders trained autoregressively to generate a velocity model from shallow subsurface to deep. Owing to the fact that seismic data are often recorded on the Earth's surface, a top-down generator can utilize the inverted information in the shallow as guidance (prior) to generating the deep. To facilitate the implementation, we use an additional network to compress the velocity model. We also inject prior information, like well or structure (represented by a migration image) to generate the velocity model. Using synthetic data, we demonstrate the effectiveness of VelocityGPT as a promising approach in generative model applications for seismic velocity model building.
\end{abstract}


\section{Introduction}
Understanding the underlying data distribution is essential in many fields, such as statistics, machine learning, and data science. Some of the advantages include, but are not limited to, quantifying uncertainty, properly weighting decisions, and generating new observations \citep{tomczak2021deep}. The traditional ways of learning data distributions involve statistical techniques such as parametric and non-parametric methods. Parametric methods assume that the data follow a specific distribution characterized by a finite set of parameters, such as the normal distribution defined by its mean and variance. Examples include linear regression and Gaussian mixture models \citep{reynolds2009gaussian}. Non-parametric methods, on the other hand, do not assume a fixed form for the data distribution and can adapt to a broader range of data distribution shapes. Examples include kernel density estimation \citep{parzen1962estimation} and nearest-neighbor methods. Recently, deep learning methods have been shown to be effective tools for modeling complex data distributions, specifically through the use of generative models.

Generative models have gained significant attention in various scientific fields due to their ability to capture complex data distributions. They have been widely applied in seismic data analysis, where their ability to draw samples from the model’s learned distribution has been utilized for various tasks, including generating more observations. Examples of the applications include seismic inversion (e.g., \citealp{meng2021seismic}), as an inversion constraint (e.g., \citealp{wang2023prior}), even for uncertainty quantification (e.g., \citealp{siahkoohi2020deep}). Moreover, as has been shown by \citet{wang2024controllable}, our prior knowledge of the subsurface (e.g., from well data and seismic images) can be included in the training for the network to conditionally sample at the inference stage. A more advanced application in seismic inverse problems has been shared in \citet{taufik2024learned}, in which a diffusion model has been shown as an effective tool to store the distribution of a multi-parameter subsurface model for regularizing an elastic Full Waveform Inversion (FWI) algorithm. However, these approaches are mainly based on CNN architectures, which are tailored for images.

In the context of seismic inverse problems, which are mainly focused on inverting subsurface velocity models, the ideal situation is to have sources and receivers surrounding our target area. However, this is often economically and technically infeasible, and thus, for subsurface investigations, the industry has resorted to recording seismic data mostly on the Earth's surface. As a result, our resolution of the deep, where most of the resources reside, depends on our ability to resolve the shallow part of the Earth's subsurface. Among the famous and widely utilized techniques for velocity model building promote layer stripping \citep{yilmaz2001seismic}, which inverts the velocity model one layer, or vertical grid sample, at a time, iteratively from top to bottom. Connecting to the context of generative models, the commonly utilized generative modeling frameworks, such as flow-based models (e.g., normalizing flow \citep{rezende2015variational}) and latent variable models (e.g., generative adversarial network \citep{goodfellow2014generative}; diffusion model \citep{ho2020denoising}), are not explicitly equipped with this capability of generating the velocity models from top to bottom. On the other hand, autoregressive models inherently provide the capability of sequential prediction relying on the previous sequence as prior. In the past, we utilized CNN (e.g., PixelCNN \citep{van2016pixel}) and RNN (e.g., PixelRNN \citep{van2016pixel}) as the backbone architecture for an autoregressive framework, which has been found inefficient (for RNN) and localized (for CNN). ImageGPT \citep{chen2020generative} was introduced as a pioneering work in utilizing Transformer decoder \citep{vaswani2017attention} as the backbone architecture for autoregressive image synthesis, in which Transformer is known for its scalability and long-range entanglement. 

In this paper, we introduce a framework that opens a new avenue in generative models applied to seismic velocity model building. We call the framework VelocityGPT, which promotes the idea of generating velocity models from top to bottom using a Transformer decoder trained in an autoregressive manner. The contribution of this paper is the following:
\begin{enumerate}
    \item We introduce a novel framework that allows for naturally synthesizing seismic velocity models in a top-to-bottom manner.
    \item We show different ways of incorporating our prior knowledge of the velocity model to constrain the generation process, which includes geological features, well velocities, and seismic post-stack images.
    \item We test the proposed framework on the OpenFWI benchmarking dataset, which demonstrates the potential of VelocityGPT as a robust seismic velocity generator.
    \item We demonstrate that a VelocityGPT model trained on small velocity models could be utilized to generate velocity models on a more realistic scale without any additional training.
\end{enumerate}

The paper is structured as follows. First, we will introduce the overall framework in Section \ref{sec:theory}, which comprises of encoding the velocity models and, subsequentially, storing their distribution. The details of each step, including the training strategy, are then discussed in Section \ref{sec:results}. We will share the results and also illustrate how various conditions, including geological classes, well, and structural information (from images), can be included during training to allow for conditional sampling in the inference stage in Section \ref{sec:results}. Finally, a discussion of the results and concluding remarks are presented in Sections \ref{sec:discussions} and \ref{sec:conclusions}, respectively.

\section{Theory}
\label{sec:theory}

\subsection{The Framework}
\begin{figure}
    \centering
    \subfigure[]{\includegraphics[height=0.3\textheight]{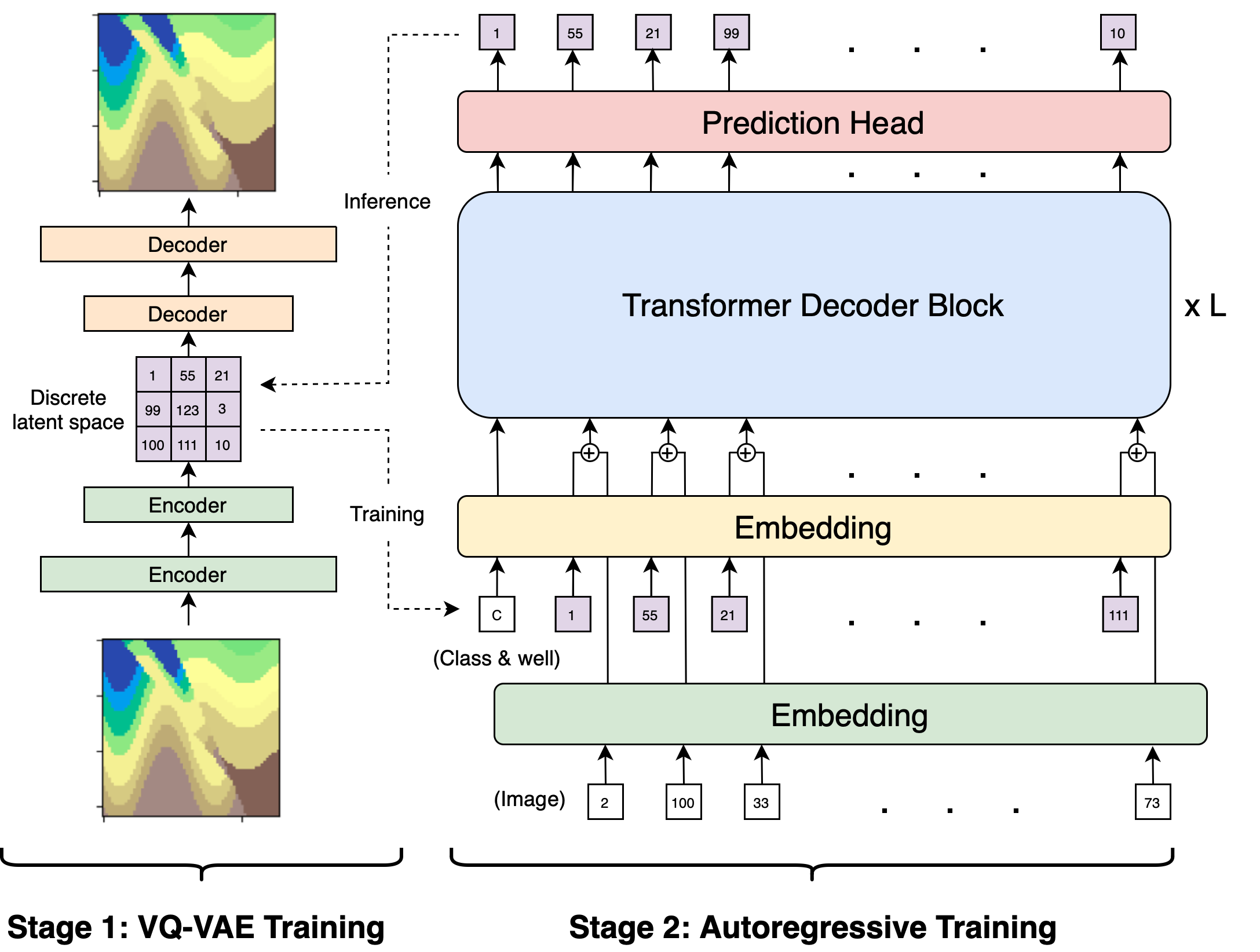}}
    \subfigure[]{\raisebox{0.135\height}
    {\includegraphics[height=0.25\textheight]{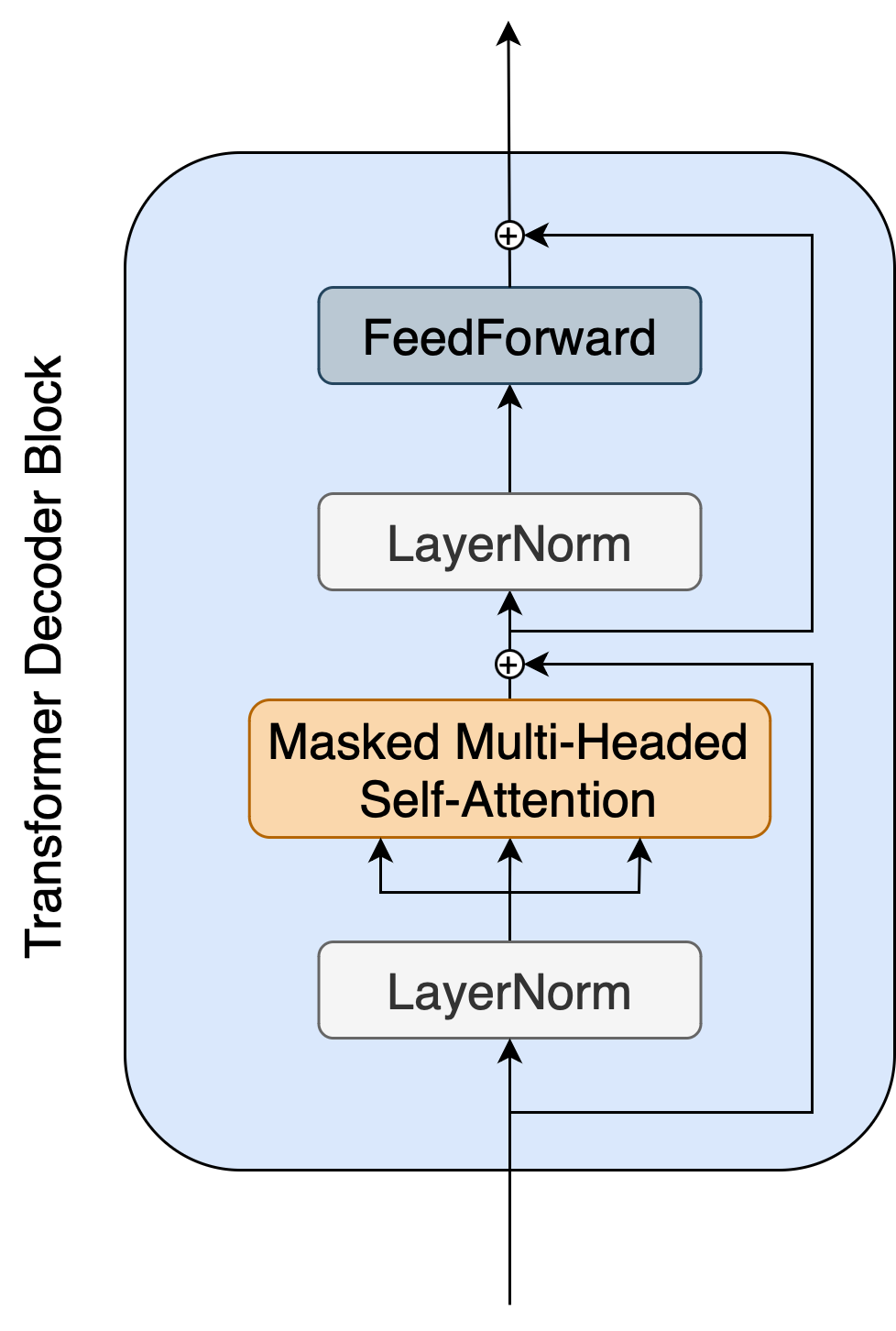}}}
    \caption{a). The framework and the network architecture of VelocityGPT. The framework consists of two stages: VQ-VAE training to convert velocity models into their discrete representation and autoregressive GPT training to model the distribution of the velocities. b). A diagram of a single Transformer decoder block.}
    \label{fig:framework-diagram}
\end{figure}

\begin{figure}[!h]
    \centering
    \includegraphics[width=0.75\textwidth]{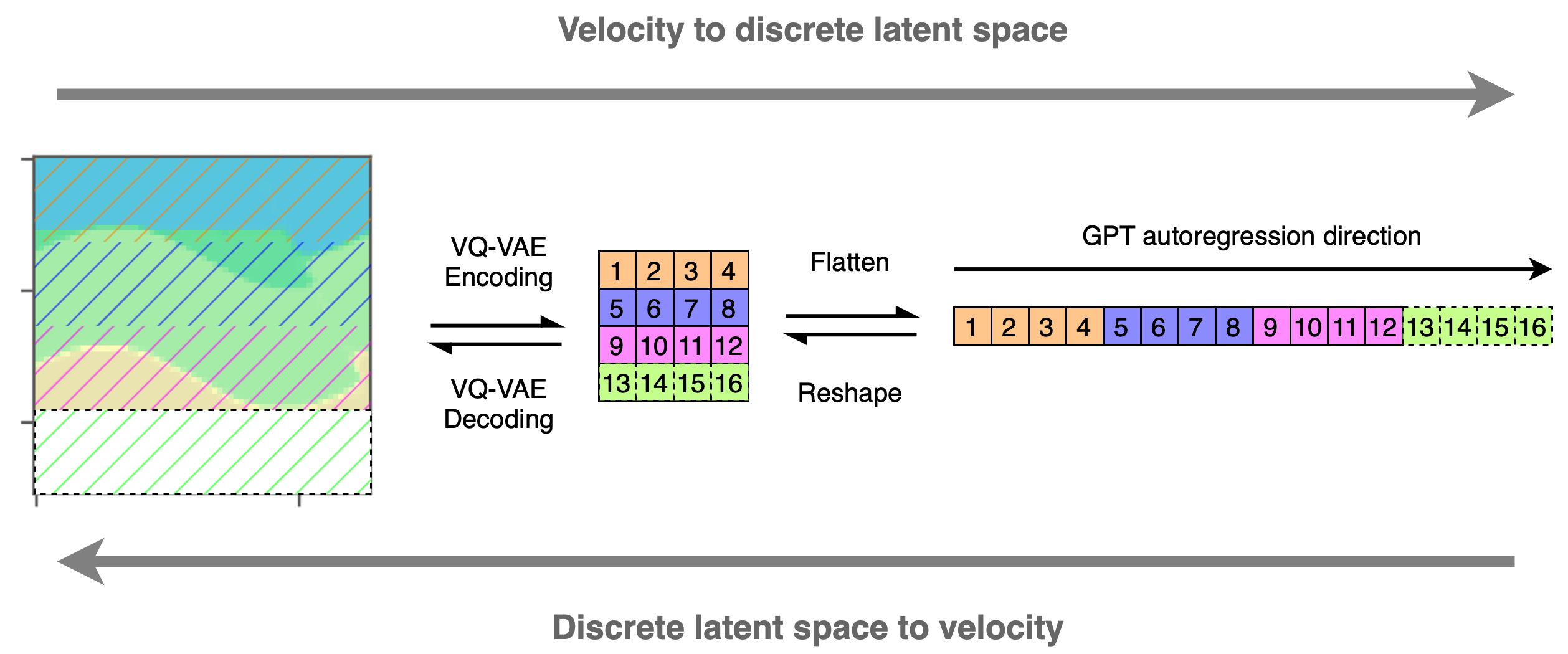}
    \caption{Velocity encoding/decoding process. Solid colors in the discrete latent domain (middle) represent the region with the corresponding shaded colors in the velocity domain (left). Dashed boxes represent the region to be generated.}
    \label{fig:encoding-diagram}
\end{figure}

The framework for our multi-conditional autoregressive velocity generator is shown in Figure \ref{fig:framework-diagram}a. In the first stage of the framework, we train a Vector-Quantized Variational Auto Encoder (VQ-VAE, \citet{van2017neural}) that transforms overlapping patches of the velocity models into their discrete forms (like classes). Then, the distribution of the velocity models is stored by training a Generative Pre-trained Transformer (GPT) model \cite[]{radford2018improving}, which is a stack of Transformer decoders with an embedding layer and an output layer, using an autoregressive objective. The top-to-bottom prediction comes naturally from how the velocity models are represented as the input to the GPT model. As illustrated in Figure \ref{fig:encoding-diagram}, a velocity model (or an incomplete one in the inference stage) is compressed (encoded) by the VQ-VAE into a discrete latent space, where each layer is represented by a set of integers. These integers are then flattened in a rasterized order (from top left to bottom right) into a long sequence of integers, which acts as one input sample to the GPT model. The GPT model predicts the next set(s) of integers that correspond to the deeper layer of the velocity model. Finally, the reverse processes of reshaping the sequence and decoding the discrete latent space with the VQ-VAE are applied to obtain the completed velocity model. The details of each stage are discussed in the following sections.

\subsection{Vector-Quantized Variational Autoencoder (VQ-VAE)}

The second stage of the framework, which involves the GPT model training, implements an autoregressive classification to generate a new velocity sample. The classification nature of the approach requires that we first convert the velocity models from continuous to discrete space. There are several approaches to obtain the discrete representation of an input. One pioneering work is introduced by \citet{van2017neural}, in which a modified autoencoder with an embedding vector in between is utilized to convert multiple modalities (audio, image, video) into their discrete representations. Recent studies build upon this concept by modifying the training or the network components (e.g., \citealt{esser2021taming}; \citealt{yu2021vector}). For simplicity, in this work, we employ the Vector-Quantized Variational Auto Encoder (VQ-VAE), which includes a convolutional autoencoder and involves an embedding implementation \cite[]{van2017neural}.

The VQ-VAE is structured as in Figure \ref{fig:framework-diagram}a, given by a two-layer encoder-decoder with an embedding matrix as a bottleneck. Each layer uses a 2D convolution with a kernel size of 4 and a stride of 2. With this configuration, the VQ-VAE has a compression ratio of $2^l$, where $l$ is the number of layers. Therefore, our VQ-VAE compresses the velocity models from 64 x 64 continuous values to 16 x 16 discrete (integer) values. A codebook size (unique discrete values) $k$ is selected to quantize the continuous space of the velocity models into a discrete space. The network is then optimized using the following objective function \cite[]{van2017neural}:
\begin{equation}
    L_{VQ-VAE} = \log p\left(x|z_d(z_e(x))\right) + ||sg[z_e(x)] - e||_2^2 \\ + \beta ||z_e(x) - sg[e]||_2^2,
\end{equation}
where $x$ is the input velocity model, $z_e$ is the encoder, $z_d$ is the decoder, $e$ is the embedding (discrete space), and $sg$ is a stop gradient operator defined as identity in the forward pass and zeros in the backward pass. The role of each term in the objective function is as follows: The first term is the reconstruction loss, which optimizes the encoder and decoder; The second term is the embedding loss, which updates the dictionary; Finally, the third term is a commitment loss, which constrains the encoder and the embedding.

\subsection{Generative Pre-trained Transformer}
For the autoregressive training, we are inspired by the approach of ImageGPT \cite[]{chen2020generative}, in which a pixel is treated as a token, and the objective is to predict the next pixel (or its quantized version) given the previous pixels (or their quantized versions). ImageGPT is based on the Transformer decoder \citep{vaswani2017attention}. We adopt the ImageGPT model and introduce modifications to optimally adapt VelocityGPT to the input of discrete latent codes of velocity models (instead of pixels) along with their conditions. The model consists of three main components: the \textit{embedding block}, \textit{decoder block(s)}, and a \textit{prediction head}. One-hot vectors of the encoded velocity model $e\;\in\;\{0, 1\}^{S \times k} \colon \sum_{i=1}^k e_i = 1$, which is the output of the trained VQ-VAE, acts as our input, where $S$ is the length of the flattened, one-hot-encoded discrete latent representation of the velocity model and $k$ is the velocity codebook size. In the embedding block, we use an embedding layer $E_1$ that projects every one-hot-encoded discrete latent code to a hidden dimension $H$:
\begin{equation}
    \label{eq:embedding-1}
    E_1(e) = e W_{E_1} + b_{E_1},
\end{equation}
where $W_{E_1} \in \mathbb{R}^{k \times H}$ and $b_{E_1} \in \mathbb{R}^{H}$ are the weights and biases of the embedding layer, respectively. Then, a learnable positional embedding $E_2$ is added to the linearly projected data:
\begin{equation}
    \label{eq:embedding-2}
    E_2(r) = r W_{E_2} + b_{E_2}
\end{equation}
where $r \in \mathbb{Z}^S$ is the relative position of the discrete latent codes, $W_{E_2} \in \mathbb{R}^{S \times H}$ and $b_{E_2} \in \mathbb{R}^{H}$ are the weights and biases of the positional embedding layer, respectively. Therefore, the output of the embedding block is:
\begin{equation}
    \label{eq:embedding-all}
  \hat{E} = E_1(e) + E_2(r).  
\end{equation}

Next, the embedded velocity model $\hat{E}$ is forwarded to a stack of $L$ decoder blocks. The diagram of a decoder block is illustrated in Figure \ref{fig:framework-diagram}b, which is composed of a masked multi-headed self-attention (MMSA) layer, a position-wise feed-forward (FFN) layer, and layer normalization along with residual connections. Inside the MMSA layer, the embedded velocity model $\hat{E}$ is first linearly projected into three different matrices, namely query ($Q$), key ($K$), and value ($V$):
\begin{equation}
    \label{eq:qkv}
    Q = \hat{E} W_Q + b_Q,\;\;\; K = \hat{E} W_K + b_K,\;\;\;V = \hat{E} W_V + b_V,
\end{equation}
where $W \in \mathbb{R}^{H \times H}$ and $b \in \mathbb{R}^{H}$ are learned weights and biases, respectively, of the linear projection for the query, key, and value matrices, denoted by their respective symbols as subscripts. Then, the masked attention 
($MA$) operation is applied to $Q$, $K$, and $V$, which is defined as:
\begin{equation}
    \label{eq:mmsa}
    X = MA(Q, K, V) = \text{softmax}\left(\frac{QK^T + M}{\sqrt{H}}\right)V,
\end{equation}
 where $M \in \mathbb{R}^{S \times S}$ is a mask matrix containing zeros on the diagonal and the lower triangular and negative infinity elsewhere (upper triangular). This mask enforces the attention maps to be a lower triangular matrix, thus preventing the model from utilizing the future information (next velocity layers) in the input sequence. The relationship (weights) between the discrete latent codes of the velocity model is stored in the query-key dot product. These operations are distributed over the number of attention heads $A$, which are done by splitting the input beforehand and concatenating them back afterwards.

Matrix $X$ is then added with a residual connection and normalized with a layer normalization. Further, this matrix is passed to a feed-forward network (FFN). The FFN is a simple two-layered fully-connected neural network with a Gaussian Error Linear Unit (GELU) activation function in between \citep{vaswani2017attention, devlin2018bert}. The output of the FFN layer is denoted as:
\begin{equation}
    \label{eq:ffn}
    F = \psi_1(X W_1 + b_1)W_2 + b_2.
\end{equation}

where $\psi_1$ is a GELU function, $W$ are the weights, and $b$ are the biases. Once again, a residual connection is applied to $F$. Finally, the output of the final encoder block goes into the prediction head $P$. The role of this prediction head is to project the hidden dimension $H$ to the velocity embedding size $k$. Thus, a linear layer and a Tanh activation function $\psi_2$ are used as follows:
\begin{equation}
    \label{eq:prediction-head}
    P(F) = \psi_2(F W_P + b_P),
\end{equation}

The original ImageGPT promotes a pixel-wise prediction to generate an image. However, instead of predicting pixels, we predict a discrete vector that represents a patch from a velocity model (encoded by the VQ-VAE) based on the previous sequence of discrete vectors, in which the input probability distribution can be formulated as:
\begin{equation}
    p(e) = \prod_{i=1}^{n} p(e_{\pi_i} | c, e_{\pi_1}, \ldots, e_{\pi_{i-1}}, \theta),
\end{equation}
where $\pi_i$ is the position of the discrete vectors $e$ (obtained from the output of VQ-VAE encoder) flattened in a rasterized order (from top left to bottom right), and $c$ are the conditions on the velocity model applied to the generator. Therefore, a cross-entropy loss ($CE$) is used to optimize the model:
\begin{equation}
    CE(p, q) = -\sum_i^k p(e) \text{log} q(e),
\end{equation}
where $q(e)$ is the predicted probability distribution of the velocity models (i.e., the output of the VelocityGPT model). Equation 9 allows for a straightforward implementation of the conditions, which can be incorporated through prepending the sequence of discrete vectors of a velocity model with the corresponding conditions. As shown by \citet{zhan2022auto}, multiple image conditions (e.g., sketch, text, audio) are prepended to an input of discrete latent representation of an image, creating a long vector. In this work, to impose the classes, we create a new embedding that converts the 10 classes of the velocity models into a latent space and prepends this to the embedded discrete vectors. For injecting well information, during training, we randomly pick a vertical profile from each velocity model and use its discrete representation, given by 16 discrete elements. Then, using another embedding layer, we project these 16 discrete elements along with the well location into a latent space and prepend them to the input sequence. 

A similar approach can be applied to incorporate the structural information from a seismic image. To obtain the image of each velocity model, we utilize the 1D post-stack linear modeling operator:

\begin{equation}
    d(z) = \frac{1}{2} s(z) \circledast \frac{\text{d} \; \text{ln} \; I(z)}{\text{d}z},
    \label{eq:post-stack}
\end{equation}

where $d$, $s$, and $I$ represent the post-stack data, wavelet, and acoustic impedance in depth, respectively. The acoustic impedance is formally defined as:

\begin{equation}
    I(z) = V(z) \; \rho(z),
\end{equation}

where $V$ and $\rho$ are the velocity and density models in depth, respectively. We estimated the density from the velocity through Gardner's equation \citep{gardner1974formation}. Then, to compress the rich structural information in the image, we use another VQ-VAE with a similar architecture as the one for the velocity models.

However, since the image has the same dimension as the velocity model, prepending the image directly to the sequence means introducing another 256 (16 x 16) elements to the input sequence. A long sequence is undesirable for our Transformer decoder model as the computational cost will grow quadratically as a function of the input sequence length. Therefore, as illustrated in Figure \ref{fig:framework-diagram}a, we instead add the latent representations of the image (projected via an image embedding) to their corresponding location in the actual space along with the latent representations of the velocity model.

\section{Numerical Experiments}
\label{sec:results}

\subsection{Dataset}
We use the OpenFWI dataset \cite[]{deng2022openfwi}, accessible publicly, which contains a diverse set of velocity models with their corresponding shot gathers. For our purposes, we only extract the velocity models, which are categorized into 8 classes that represent different geological features and complexity: FlatVelA, FlatVelB, CurveVelA, CurveVelB, FlatFaultA, FlatFaultB, CurveFaultA, CurveFaultB (Figure \ref{fig:unconditional}). Each model is cropped to 64 x 64 from the original size of 70 x 70, where each pixel has a value ranging from 1500 to 4500 m/s. Additionally, we create two new velocity classes: CurveVelS and FlatVelHR. The first is obtained by smoothing the CurveVel class, representing smooth velocity models (e.g., as initial models for inversion purposes). The latter is obtained by combining and compressing four randomly picked velocity models from the FlatVel class into one sample, representing high-resolution velocity models (e.g., for reservoir modeling purposes). Finally, a total of 381k velocity samples are obtained, which are then split into 324k and 57k samples for training and validation, respectively. We also normalize the velocities into [-1, 1].

For the images, we assume a typical post-stack image and choose a Ricker wavelet with a peak frequency of 20 Hz and a sampling rate of 16 ms as a source wavelet. Then, using Equations 11 and 12, we can obtain the synthetic seismic images corresponding to each velocity model. Examples of the images are shown in the first column in Figure \ref{fig:vqvae-img}.

\subsection{VQ-VAE Training}
The architecture of the VQ-VAE is as mentioned in Section \ref{sec:theory}, which consists of four convolutional layers (two encoders and two decoders) with an embedding layer in the middle. We prepare two VQ-VAE models; one encodes the velocity models ($\text{VQ-VAE}_v$), and the other encodes the post-stack images ($\text{VQ-VAE}_i$). For $\text{VQ-VAE}_v$, we use an embedding size of 128, 128 channels in the input and output convolution layers, and 68 channels in the intermediate convolution layers, resulting in a total of 440,641 parameters. Meanwhile, for $\text{VQ-VAE}_i$, we use an embedding size of 128 and 32 channels in all the convolution modules, resulting in a total of 79,841 parameters. 

We picked the L2 loss as the reconstruction loss with $\beta = 0.1$. For other training parameters of the $\text{VQ-VAE}_v$, we use a batch size of 128 and an Adam optimizer with a learning rate of 1e-3. For the $\text{VQ-VAE}_i$, we use a batch size of 1,024 and an Adam optimizer with a learning rate of 4e-3. An early stopping module with a patience of 10 epochs monitors the training and stops whenever the validation loss does not decrease within the patience period. Under these settings and using a single NVIDIA RTX 8000 GPU, the models were updated for 142 epochs (8.3 hr) and 82 epochs (1 hr) for $\text{VQ-VAE}_v$ and $\text{VQ-VAE}_i$, respectively.

\begin{figure}[!h]
    \centering
    \subfigure[]{\includegraphics[width=0.75\textwidth]{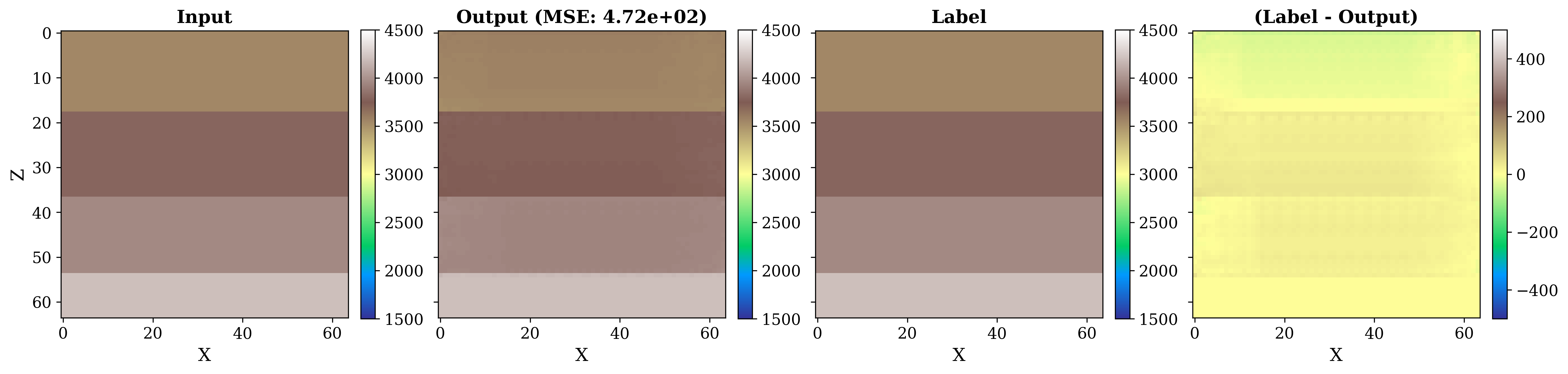}}
    \subfigure[]{\includegraphics[width=0.75\textwidth]{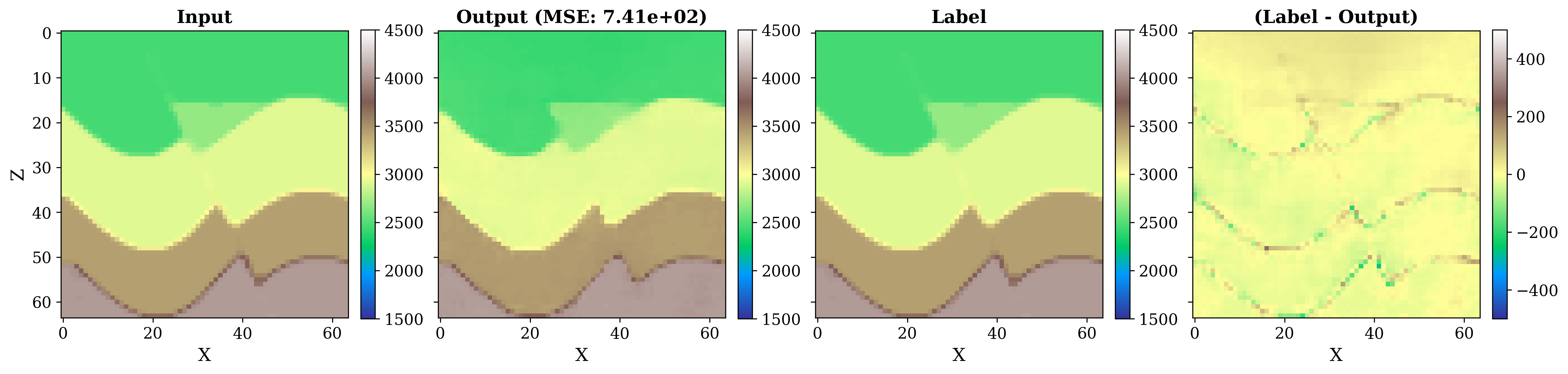}}
    \caption{Two examples of VQ-VAE post-training velocity reconstruction, which corresponds to FlatVelA (a) and CurveFaultA (b) classes.}
    \label{fig:vqvae-vel}
\end{figure}

\begin{figure}[!h]
    \centering
    \subfigure[]{\includegraphics[width=0.75\textwidth]{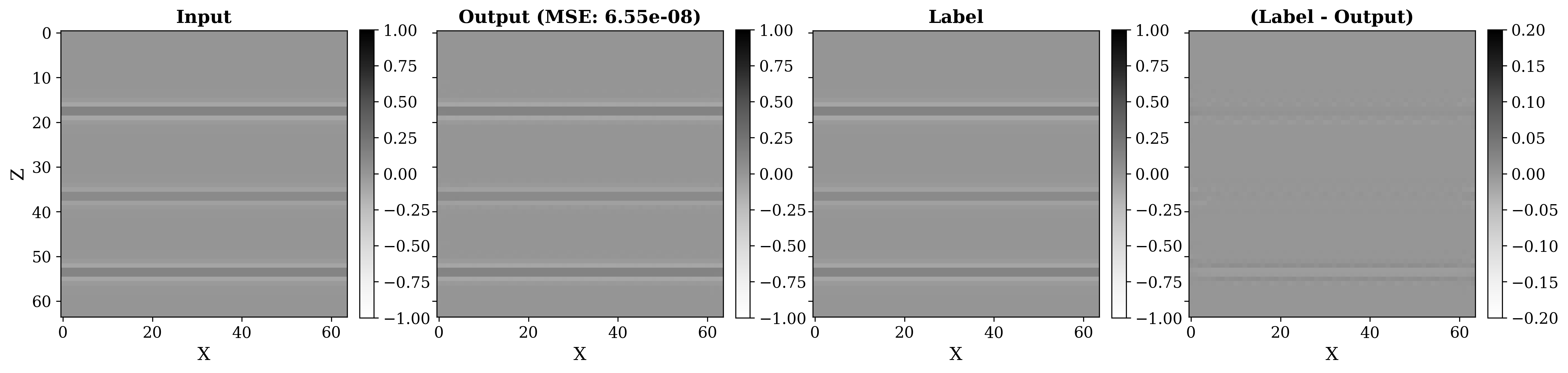}}
    \subfigure[]{\includegraphics[width=0.75\textwidth]{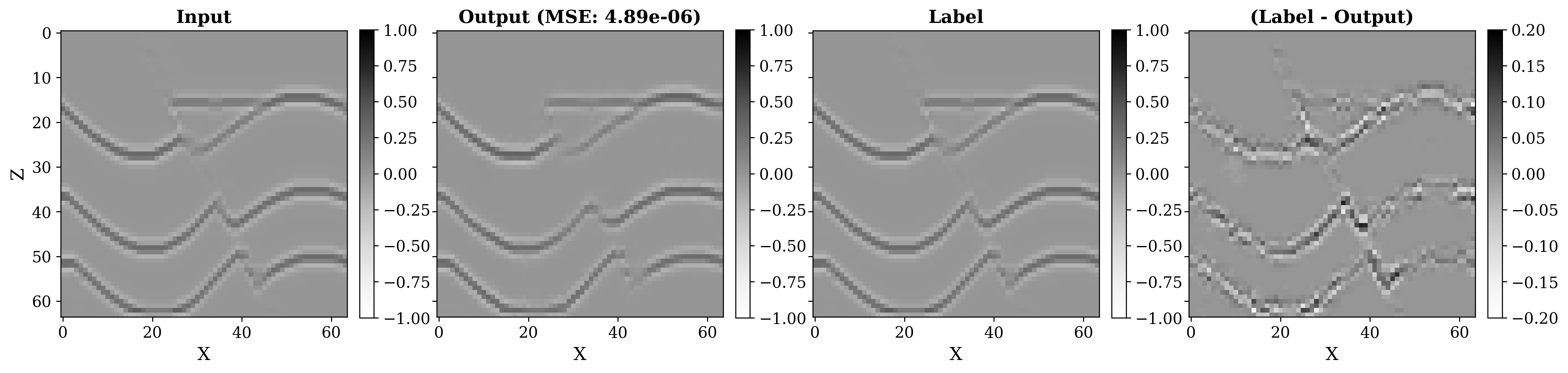}}
    \caption{Two examples of VQ-VAE post-training post-stack image reconstruction, which corresponds to FlatVelA (a) and CurveFaultA (b) classes.}
    \label{fig:vqvae-img}
\end{figure}

Figure \ref{fig:vqvae-vel} shows the reconstruction result of $\text{VQ-VAE}_v$ on two examples of the velocity model classes. The model managed to reconstruct the velocity models with minimum error (Figure \ref{fig:vqvae-vel}, rightmost column), where the loss is only visible mostly at the interfaces due to abrupt changes in the velocities. A similar observation is depicted in two examples in Figure \ref{fig:vqvae-img} for the image reconstruction of $\text{VQ-VAE}_i$. Specifically, in Figure \ref{fig:vqvae-img}, the model slightly struggles in resolving extreme structures, but the overall image is well resolved.

\subsection{GPT Training}
The standard symbols used to describe a configuration of a Transformer architecture are $L$ (the number of decoder layers), $H$ (the size of hidden/latent dimensions), and $A$ (the number of attention heads). For our tests, the GPT model is configured with $L = 8$, $H = 128$, and $A = 4$, resulting in a total of 1,680,768 trainable parameters. The pre-layer-normalization variant of the Transformer decoder is used \citep{xiong2020layer}, where the layer normalization modules are placed before the self-attention and feed-forward modules, as illustrated in Figure \ref{fig:framework-diagram}b.

The network is trained using a cross-entropy loss between the probability distributions of the output and the input discrete vectors of the velocity models (Equation 10), which is essentially self-supervised training. We employed the \textit{teacher-forcing} training strategy, where the labels are treated as the input to the network, as has been commonly done in the training of modern autoregressive models. The network parameters are optimized using an Adam optimizer with a learning rate of 4e-3 and a batch size of 1,024. With an early stopping module, the training lasts for 102 epochs, which roughly took 10.5 hours on a single NVIDIA RTX 8000 GPU.

\subsection{Inference}
In this section, we will test the capability of VelocityGPT to generate new samples of velocity models. We will start with an unconditional sampling test, in which the network is free to generate velocity models. Later, we progressively add different conditions (class, well, and image) to constrain the generation process and observe how the network handles these various conditions.

\begin{figure}[!h]
    \centering
    \includegraphics[width=0.7\textwidth]{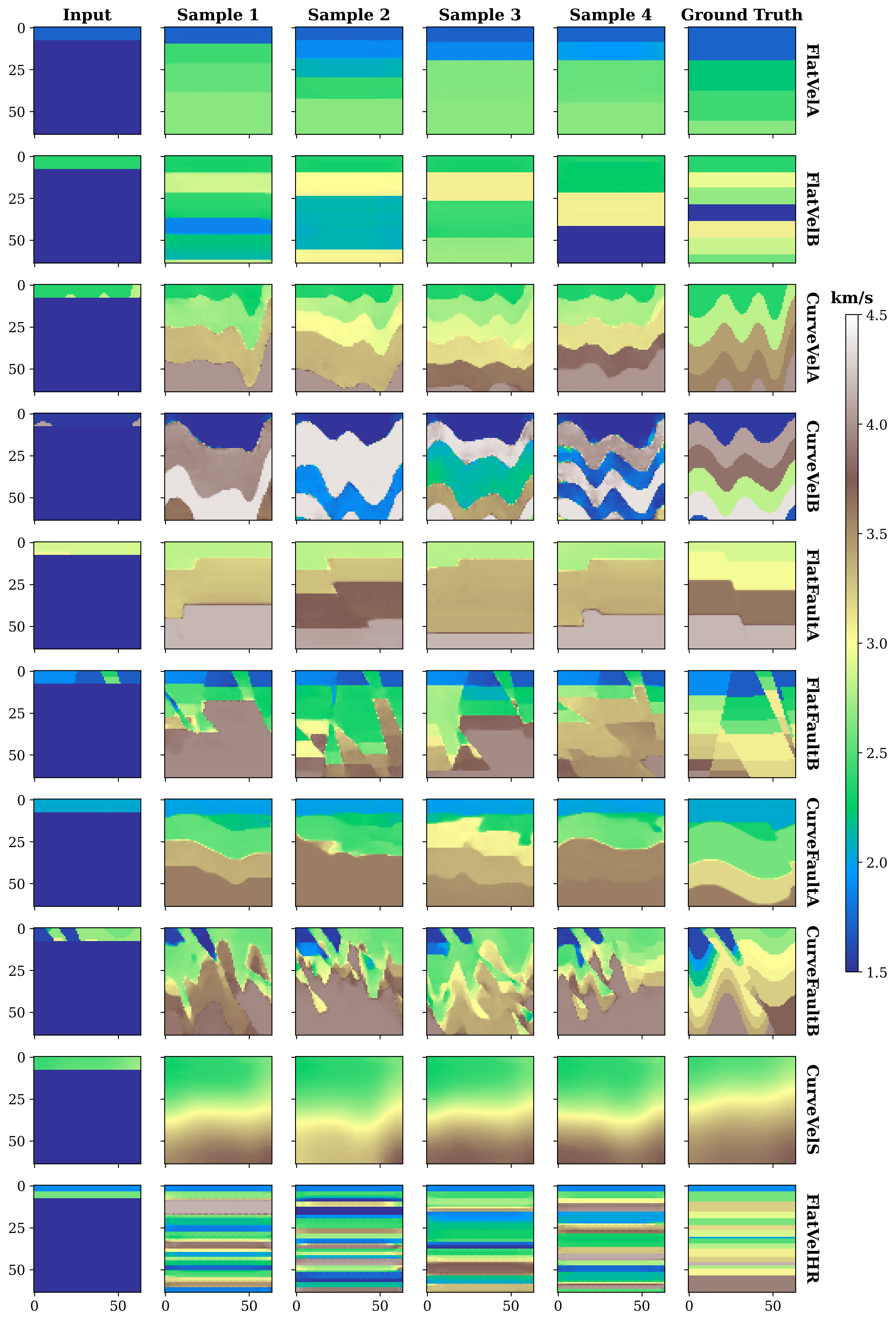}
    \caption{Samples from every class, given the input from the shallow part of a randomly picked model from the validation set displayed on the right column.}
    \label{fig:unconditional}
\end{figure}

We first test the unconditional sampling capability of the trained network by only feeding the network with shallow subsurface information as a starting point and letting the network generate different realizations based on the input. As shown in Figure \ref{fig:unconditional}, the network is capable of generating diverse samples of velocity models that agree with the shallow part in the input. 

\begin{figure}[!h]
    \centering
    \subfigure[]{\includegraphics[width=0.75\textwidth]{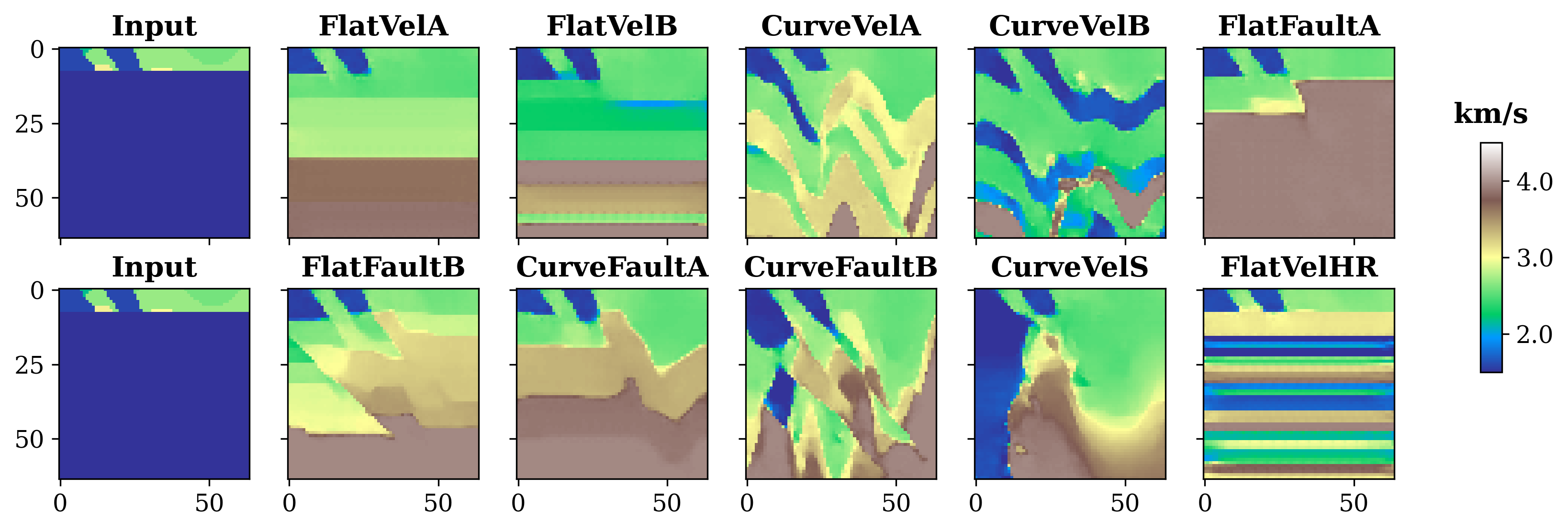}}
    \subfigure[]{\includegraphics[width=0.6\textwidth]{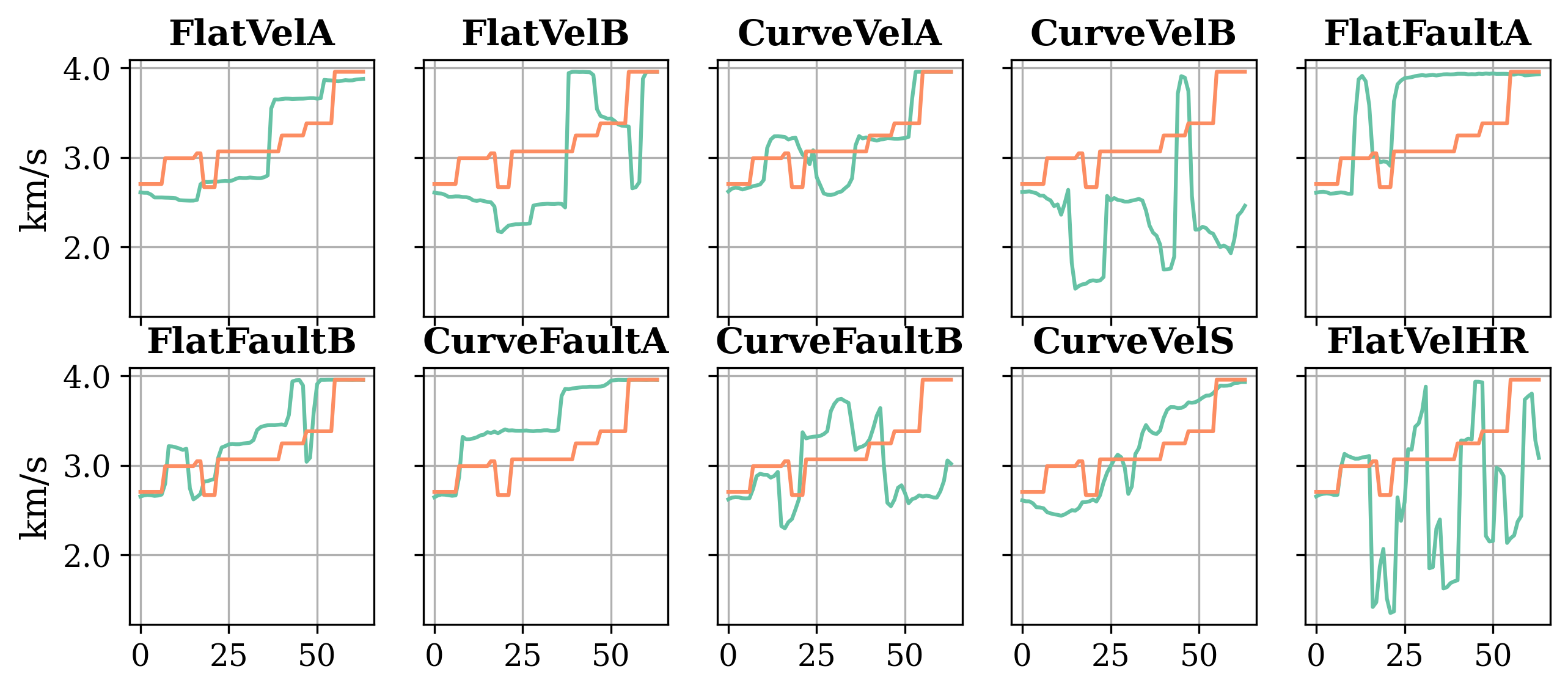}}
    \caption{a). Sampling conditioned by velocity classes, given input in the first column. The classes are labeled on the top of each subplot. b). 1D velocity profiles at X = 32, corresponding to column 1 onwards in (a). Blue and orange lines represent the generated and well velocities, respectively.}
    \label{fig:cls-cond}
\end{figure}

Next, to test the class-conditioned sampling, we fix the input shallow velocity information and vary the prepended class in the input sequence. Figure \ref{fig:cls-cond}a shows the result of the class-conditioned sampling. We can observe that the network can generate velocity models that agree with the shallow structure as well as the specified class. Note that the deeper part of the generated velocity models is not constrained (Figure \ref{fig:cls-cond}b). 

\begin{figure}[!h]
    \centering
    \subfigure[]{\includegraphics[width=0.75\textwidth]{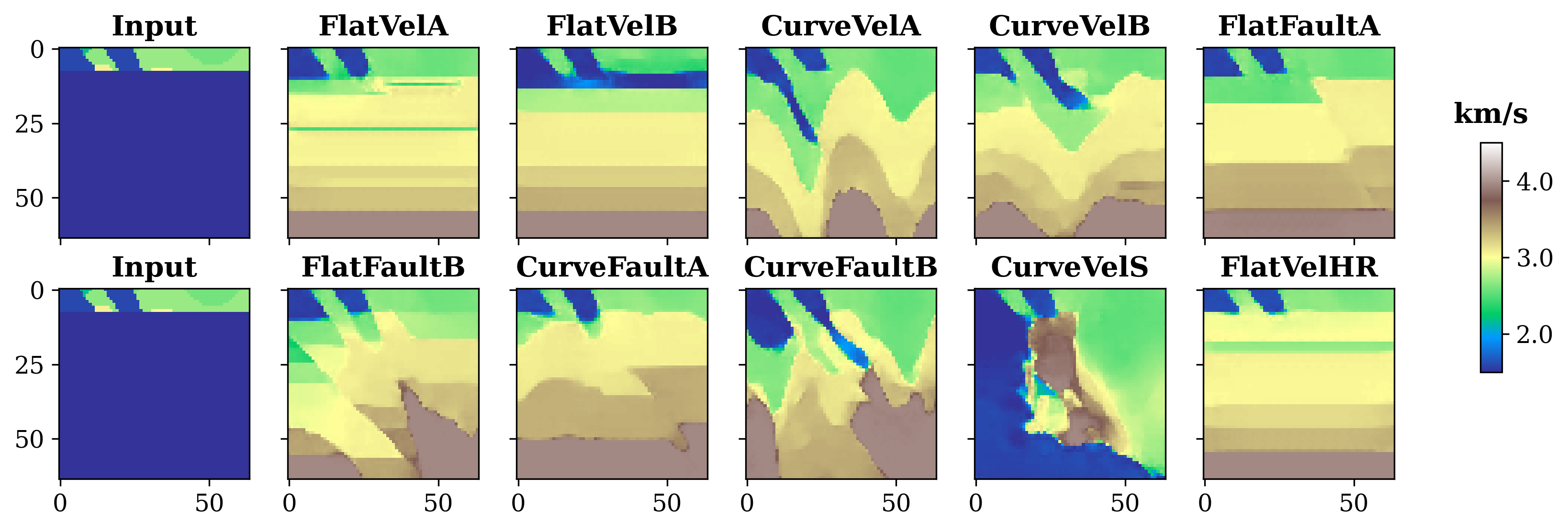}}
    \subfigure[]{\includegraphics[width=0.6\textwidth]{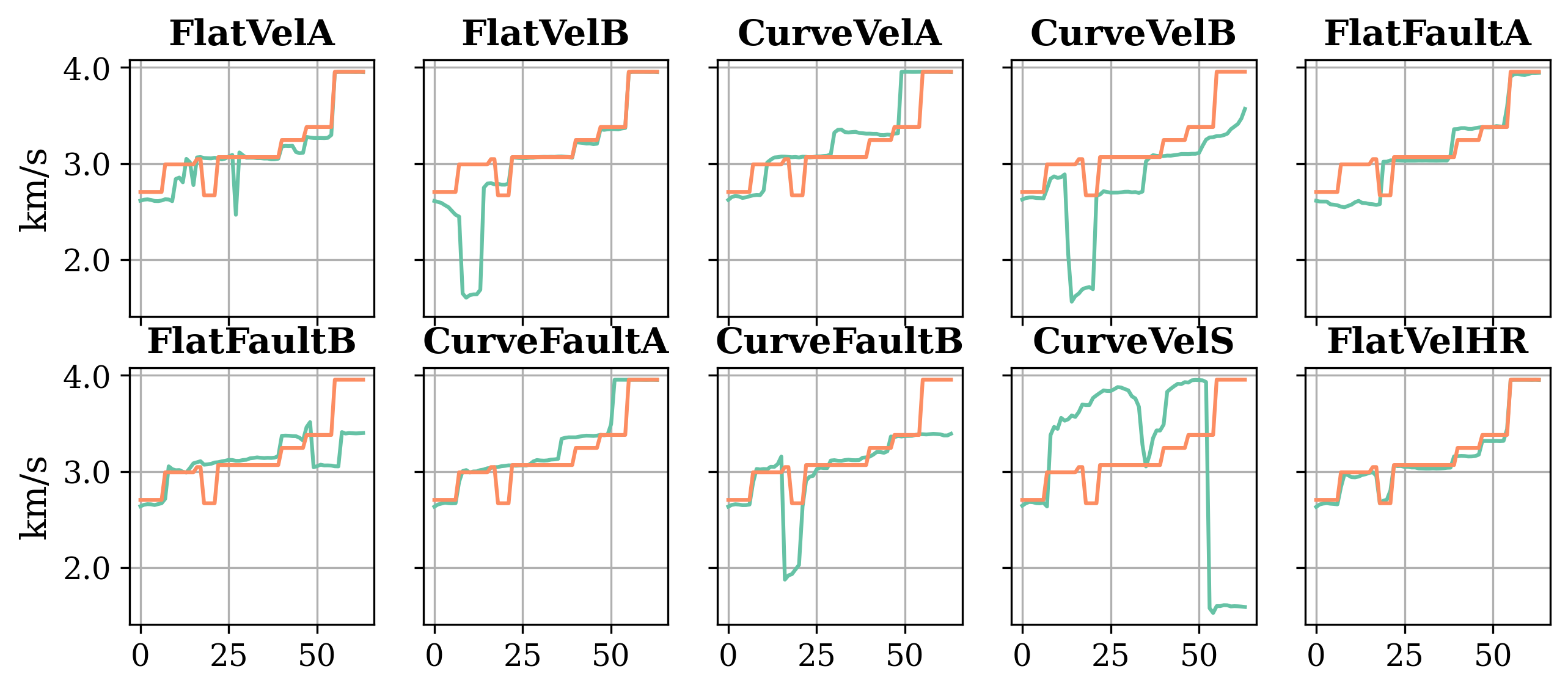}}
    \subfigure[]{\includegraphics[width=0.65\textwidth]{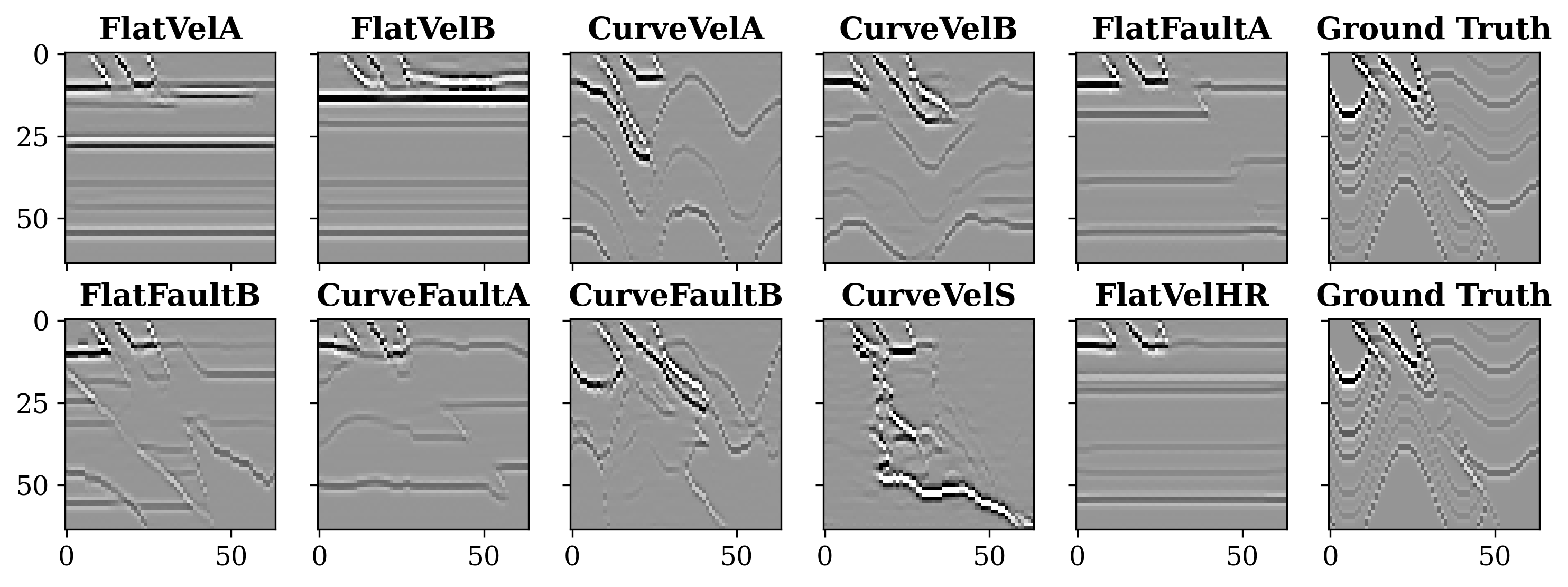}}
    \caption{a). Sampling conditioned by class and well, given input in the first column. b). 1D velocity profiles at X = 32, corresponding to column 1 onwards in (a). c). Calculated image of the generated velocities, corresponding to column 1 onwards in (a). Legends are similar to Figure \ref{fig:cls-cond}.}
    \label{fig:cls-well-cond}
\end{figure}

We then prepend the input sequence with the well profile to test the well-conditioned sampling capability of the network. As we can observe in Figures \ref{fig:cls-well-cond}a--b, the network now generates velocity models in the deeper part that agree with the well velocities, especially at the well location. This proves that the network learned to adapt to the given priors (shallow parts, classes, and wells). The structure of the model, otherwise, varies with the different samples (Figure \ref{fig:cls-well-cond}c).

\begin{figure}[!h]
    \centering
    \subfigure[]{\includegraphics[width=0.75\textwidth]{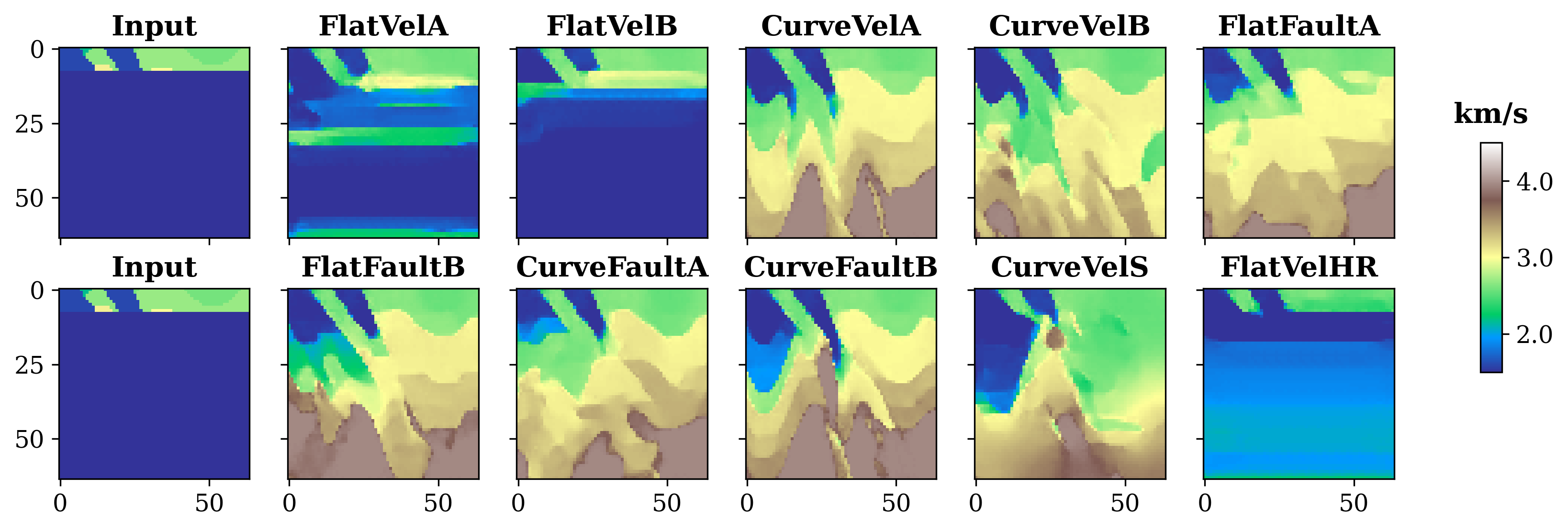}}
    \subfigure[]{\includegraphics[width=0.6\textwidth]{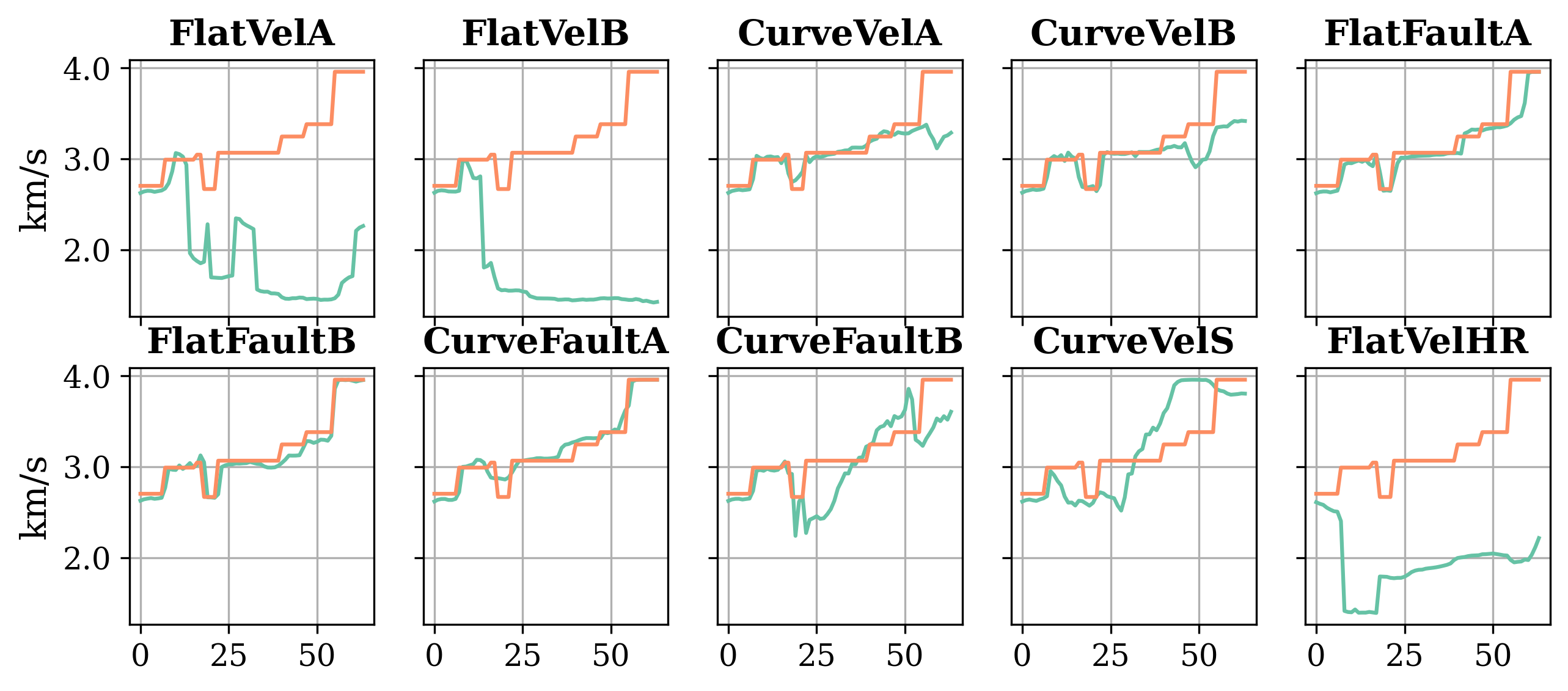}}
    \subfigure[]{\includegraphics[width=0.65\textwidth]{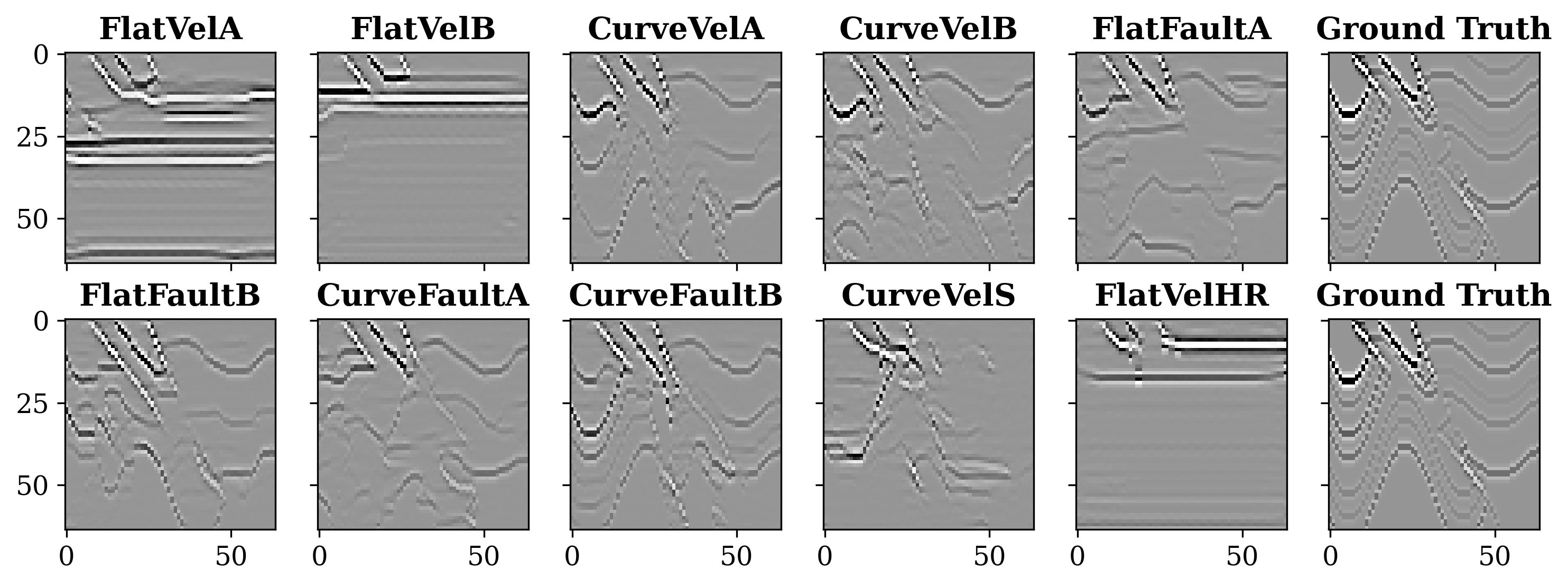}}
    \caption{a). Sampling conditioned by class, well, and image (c, rightmost column), given input in the first column. b). 1D velocity profiles at X = 32, corresponding to column 1 onwards in (a). c). Calculated image of the generated velocities, corresponding to column 1 onwards in (a). Legends are similar to Figure \ref{fig:cls-cond}.}
    \label{fig:cls-well-refl-cond}
\end{figure}

Finally, we add the image to inject structural information into our velocity model. Figure \ref{fig:cls-well-refl-cond} (rightmost column) shows a sample image we use as an input to the network. As shown in Figures \ref{fig:cls-well-refl-cond}a--c, not only the generated samples agree with the previously mentioned conditions (shallow parts, classes, and wells), but also the structure is constrained (Figure \ref{fig:cls-well-refl-cond}c). For some classes, especially classes involving flat layers, the predictions seem distorted due to conflicting information between the class and the imposed structural information; thus, the network is unable to yield geologically accurate velocity models. However, when the shallow layers, class, well, and image are all aligned, the network generated a velocity model and its corresponding image that is very close to the ground truth (compare Figure \ref{fig:cls-well-refl-cond}a 2nd row, 4th column with Figure \ref{fig:unconditional} 8th row, last column for the velocity and Figure \ref{fig:cls-well-refl-cond}c 2nd row, 3rd column with Figure \ref{fig:cls-well-refl-cond}c 2nd row, last column for the corresponding image). This shows that the model is capable of handling multiple conditions as it generates samples of velocity models.

\begin{figure}
    \centering
    \includegraphics[width=0.75\textwidth]{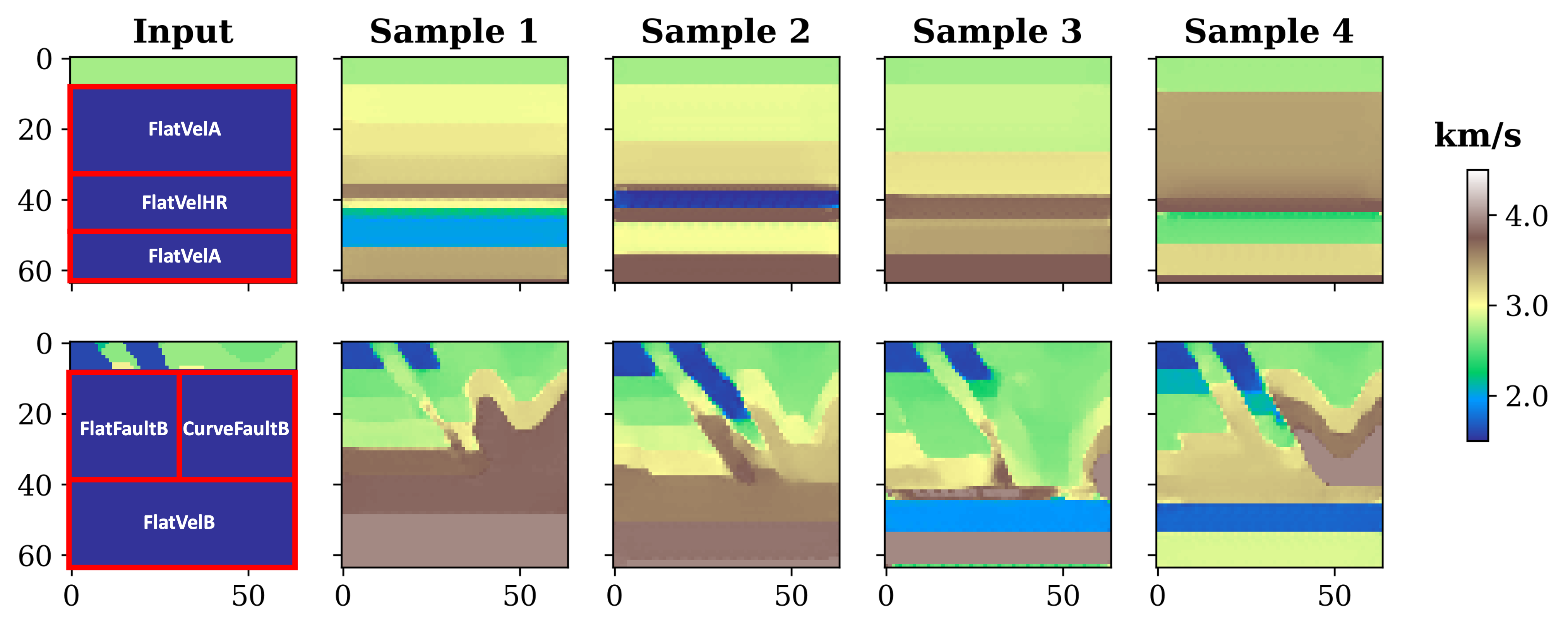}
    \caption{Generated samples conditioned spatially using classes specified in the first column.}
    \label{fig:local-pred}
\end{figure}

The sampling approach can be altered by utilizing the autoregressive nature of VelocityGPT. We can arbitrarily change the prepended class during the sequence generation process in the inference stage to force the network to generate the next discrete vectors that comply with the corresponding prepended class. Thus, by relating the sequence position to its original location in the velocity model, we can spatially control and combine different classes in the generated samples, including injecting a high resolution class (e.g., at the reservoir location), as shown in Figure \ref{fig:local-pred}.

\subsection{Scaling up to a realistic size}
In the previous section, we illustrated how to use the trained VelocityGPT to generate velocity models. However, in real scenarios, seismic velocity models typically have much larger dimensions than those we used in the training. Therefore, in the following, we share a methodology to utilize the trained VelocityGPT on a more realistic size model without the need for additional training of the network. We will also show that the conditional sampling feature of the trained network can still function well during the generation process.

\begin{figure}
    \centering
    \includegraphics[width=0.65\textwidth]{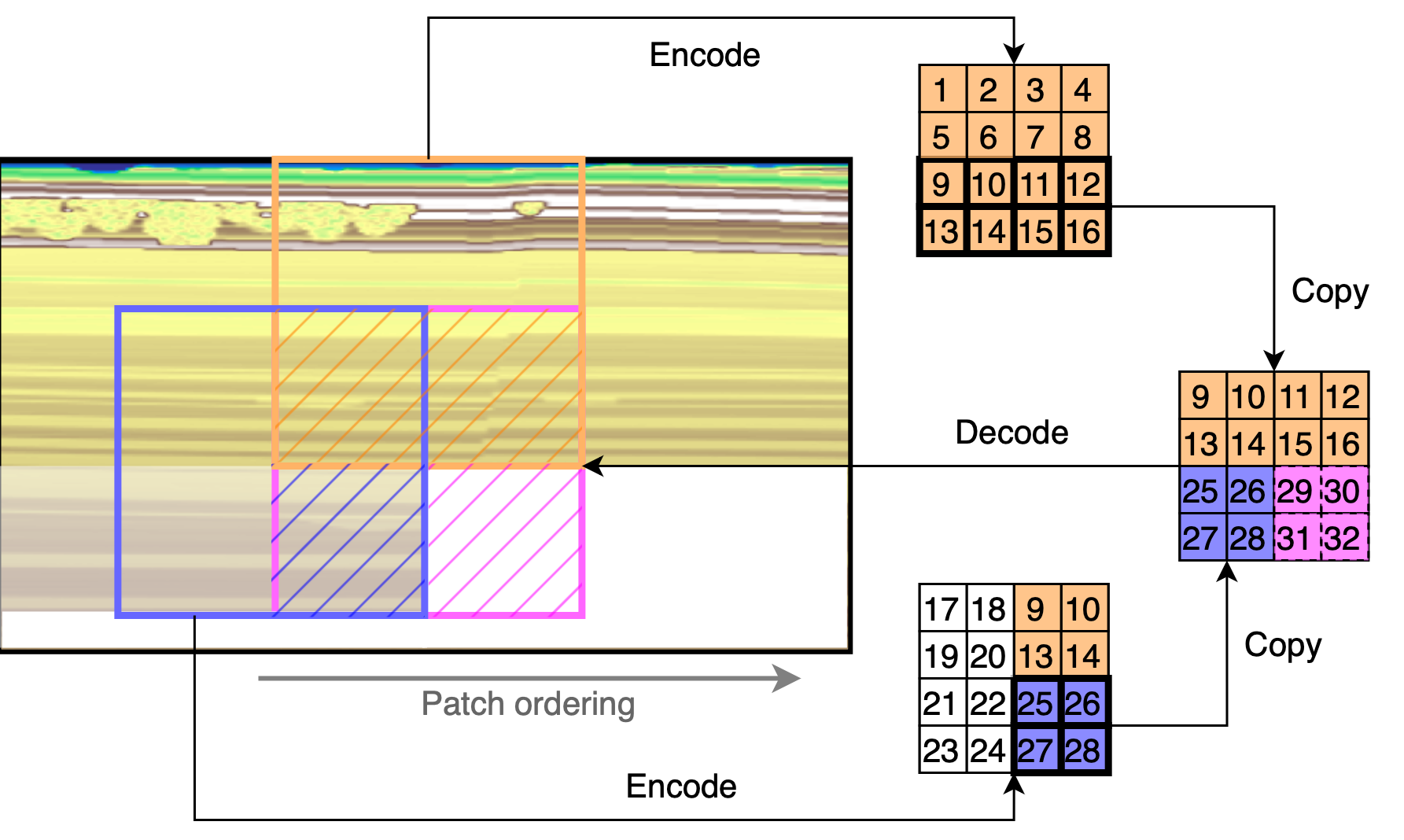}
    \caption{The inference scheme of VelocityGPT for a larger velocity model. Following the top left to bottom right ordering of the patches, the discrete latent codes of the adjacent patches are copied to the next patch according to the stride size and their actual positions in space. Then, after the discrete latent codes of the patches are completed, they are transformed back to the velocity domain. This plot is only for demonstration, but our stride is smaller than that shown in the graph.}
    \label{fig:arid-diagram}
\end{figure}

\begin{figure}[!h]
    \centering
    \subfigure[]{\includegraphics[width=0.45\textwidth]{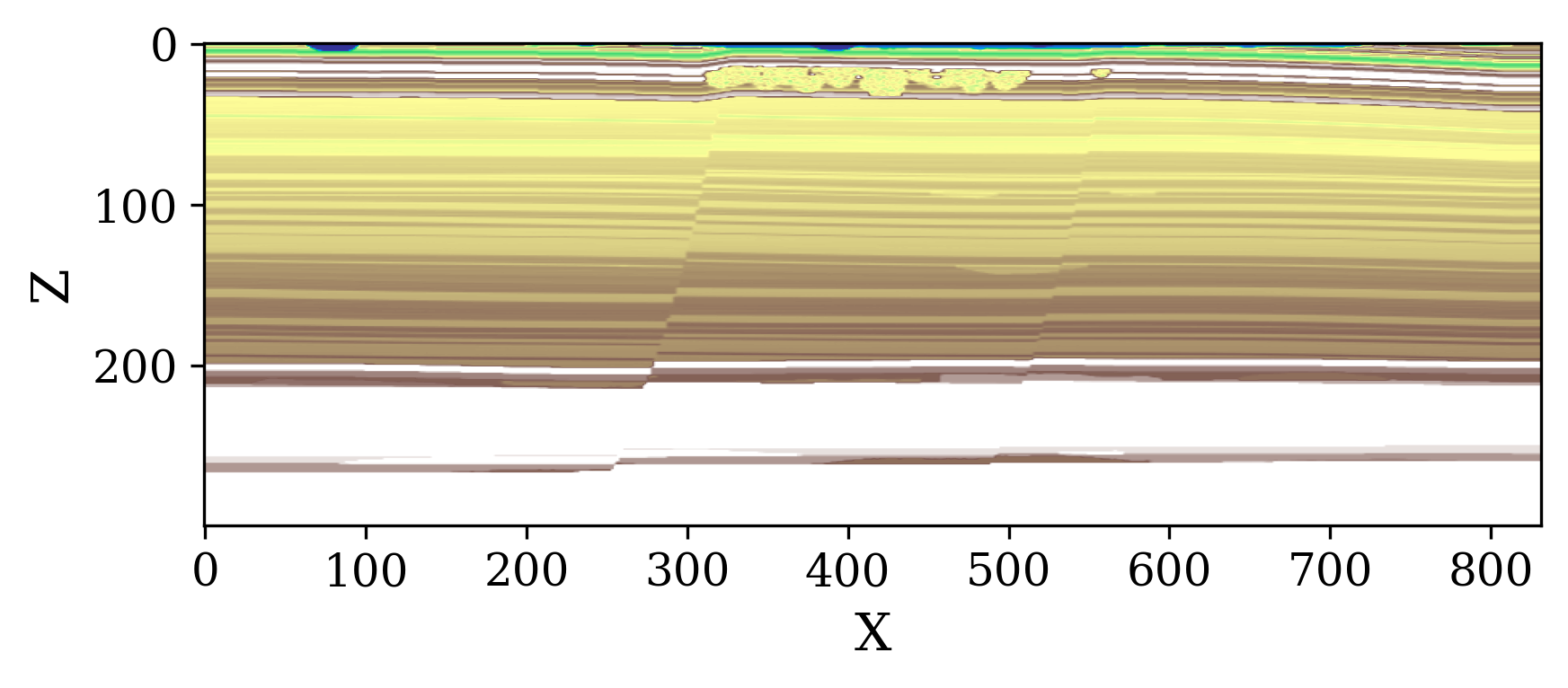}}
    \subfigure[]{\includegraphics[width=0.45\textwidth]{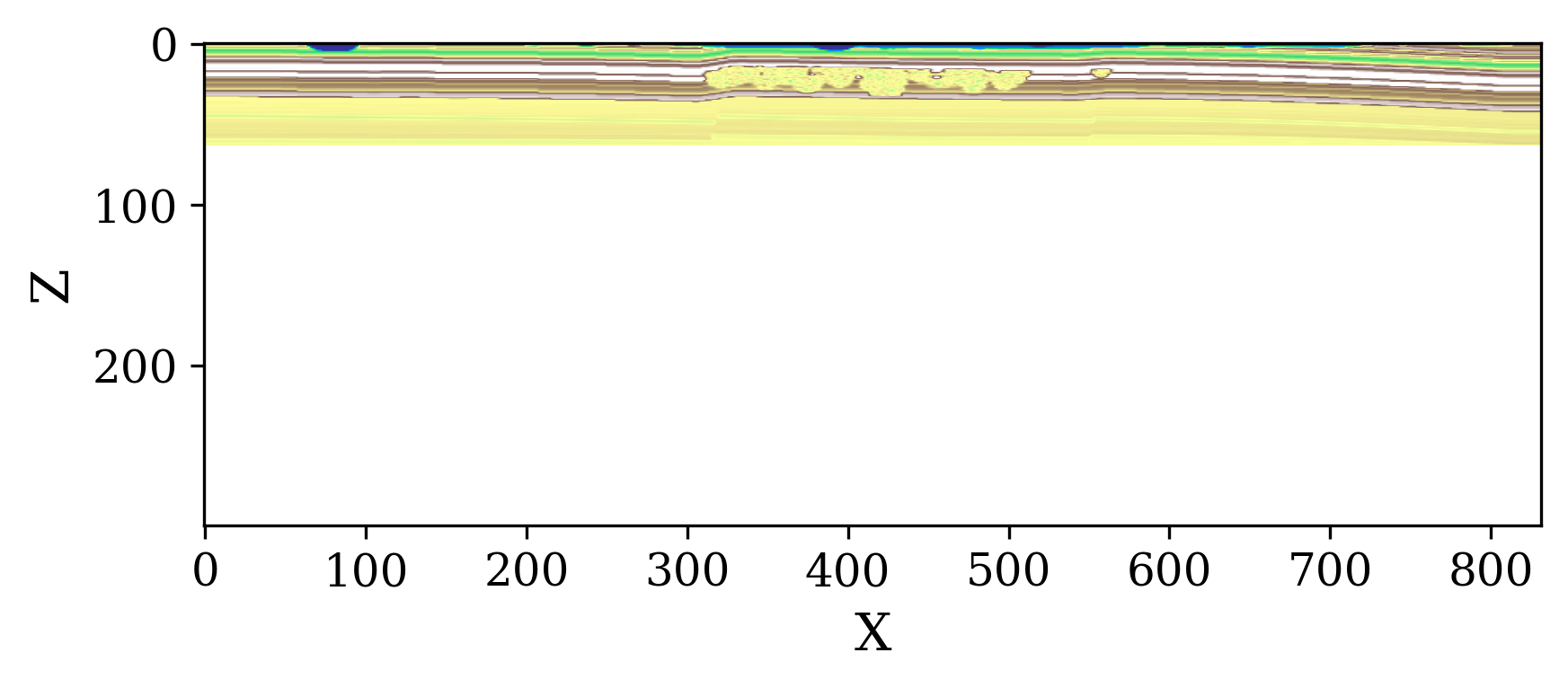}}
    \subfigure[]{\includegraphics[width=0.45\textwidth]{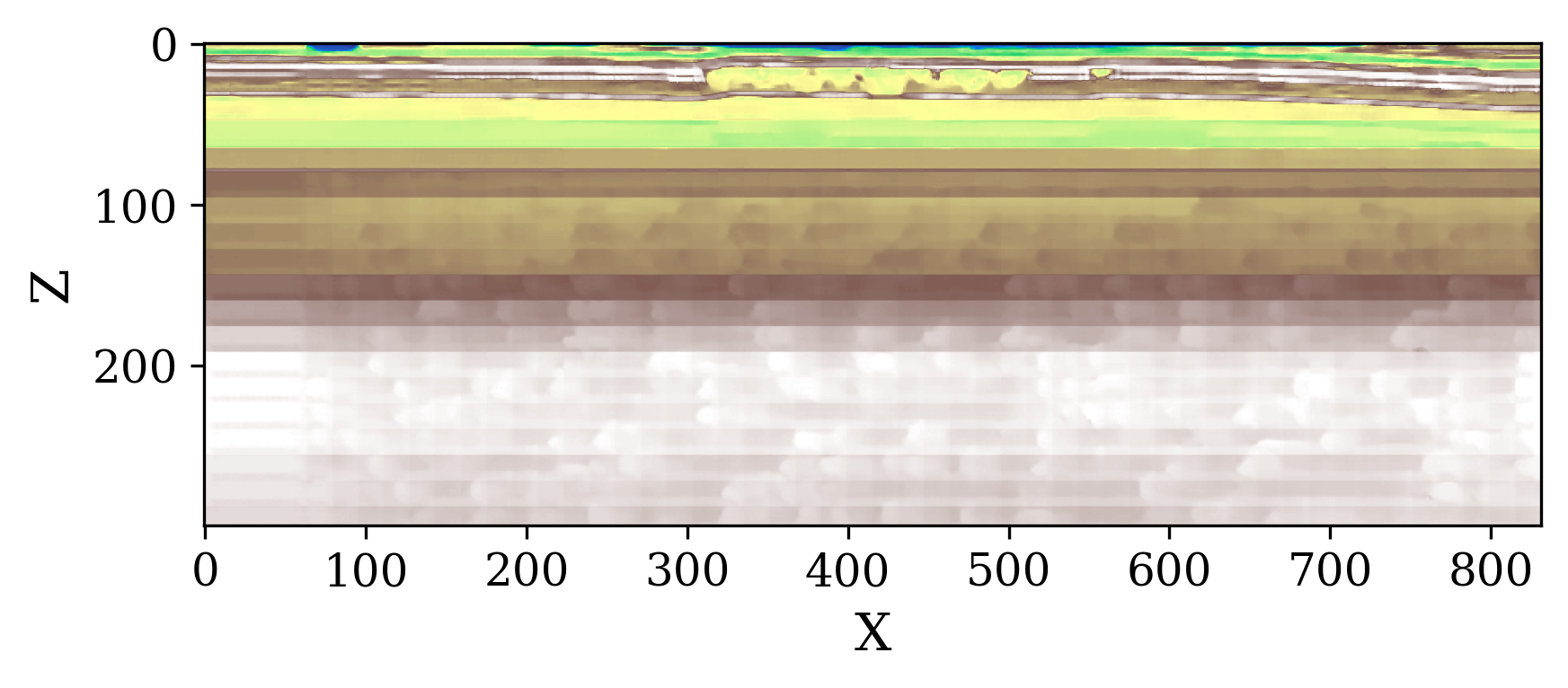}}
    \subfigure[]{\includegraphics[width=0.45\textwidth]{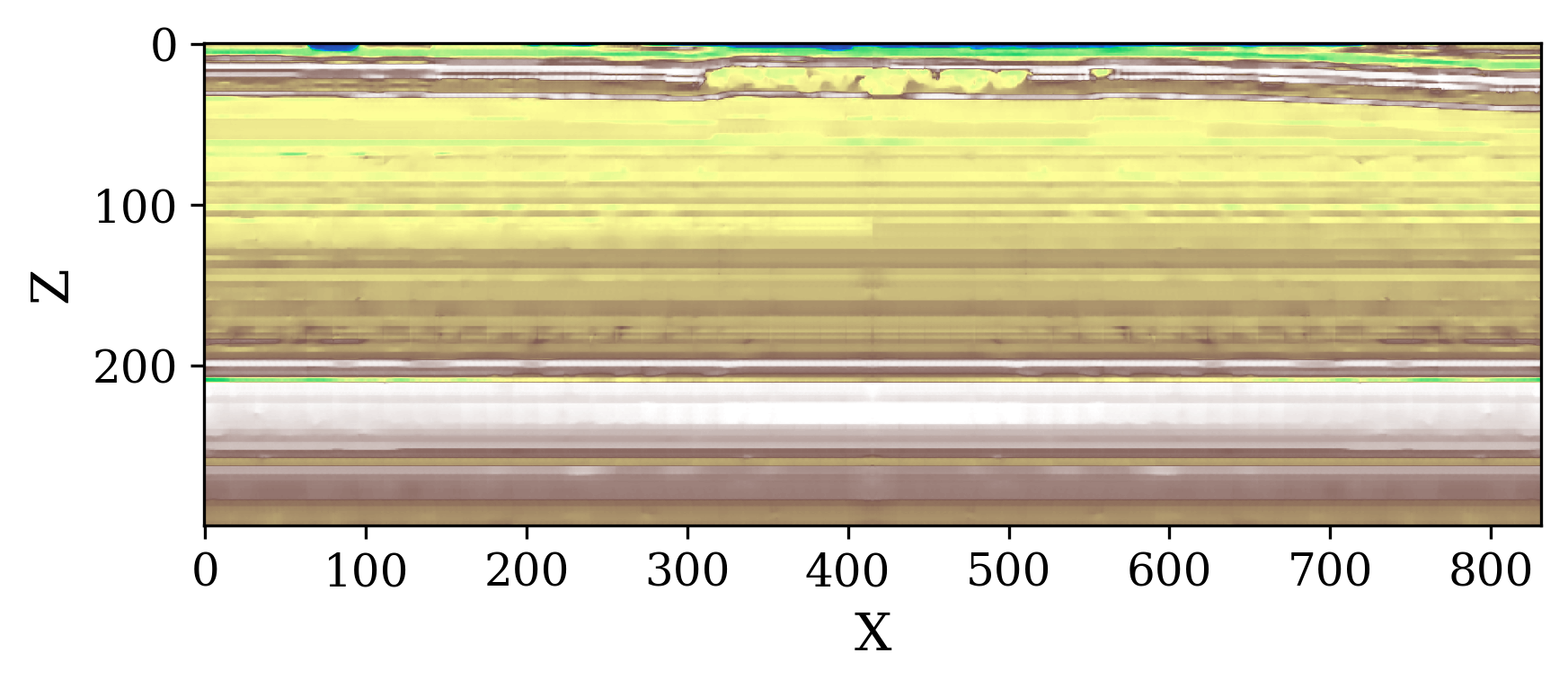}}
    \subfigure[]{\includegraphics[width=0.45\textwidth]{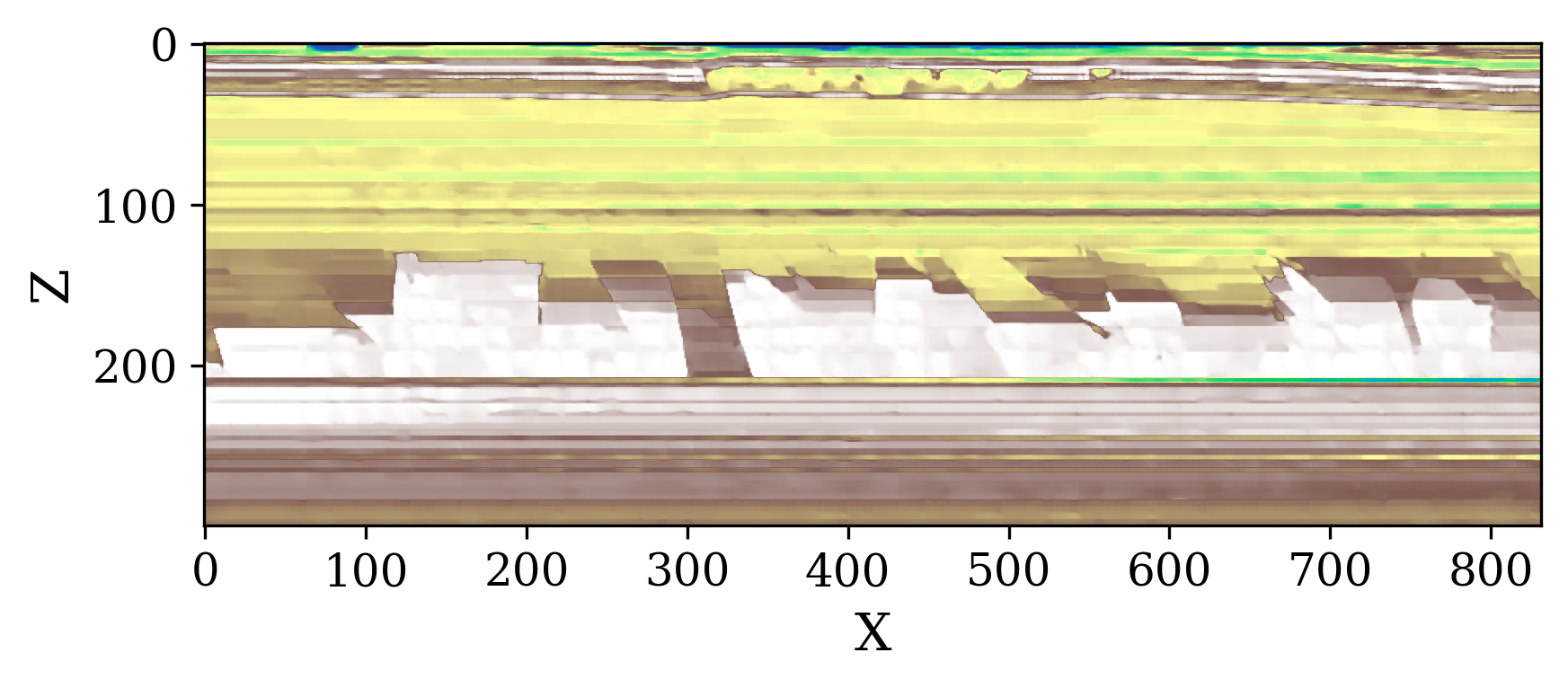}}
    \caption{a). The SEAM Arid model. b). The shallow region used as a prior. c). Generated sample conditioned with FlatVelA class. d). Generated sample conditioned with FlatVelHR class and a well located at X = 416. e). Generated sample conditioned with FlatVelHR class (top and bottom), FlatFaultB class (middle), and a well located at X = 0.}
    \label{fig:arid}
\end{figure}

To demonstrate these features, we consider SEAM Arid velocity model \citep{oristaglio2015seam} as a reference, which also serves as an out-of-distribution test. The SEAM Arid model is a complex Earth subsurface model that mimics a desert environment. As shown in Figure \ref{fig:arid}a, the model is of size 832 x 300, which is substantially larger than the OpenFWI dataset (64 x 64) that was used to train VelocityGPT. We again consider a scenario where the shallow part of the model is known (i.e., via a near-surface inversion method or a priori knowledge) and the objective is to generate the deeper part of the velocity model based on the near-surface information. Therefore, to adapt a pre-trained network to these settings, we take the following steps (illustrated in Figure \ref{fig:arid-diagram}):
\begin{enumerate}
    \item Clip the velocity values to the range of the pre-training dataset and normalize the velocities between [-1, 1] to ensure a proper encoding-decoding process;
    \item Perform overlapping patching with a specified stride to the SEAM model while preserving the top-bottom prediction scheme. Therefore, the patches are ordered from top-left to bottom-right. For our experiment on the SEAM model, some of the patches will be fully populated with prior velocity information (shallow), some we be partially (border), and most will have no information;
    \item Encode all patches using the pre-trained VQ-VAE to get the discrete latent codes of each patch;
    \item Starting from the top left, use the trained VelocityGPT model to complete the first incomplete patch we encounter in the SEAM model;
    \item Use the predicted velocity within the previous patch (in their discrete latent form)  to fill in missing information in the overlapping part of the next patch (right side) and the patch below (see Figure \ref{fig:arid-diagram});
    \item Repeat steps 4--5 until all the patches are covered;
    \item Decode the discrete latent codes independently for each patch to transform them back to patches of velocity model;
    \item Rearrange the patch to create a completed velocity model, in which overlapping regions are averaged accordingly.
\end{enumerate}

We select a patch size of 64 x 64 (the size of the OpenFWI velocity model) and a stride of 16 x 16. Moreover, we only use the top 64 samples of the SEAM Arid model as our prior and set the rest to empty (Figure \ref{fig:arid}b). Consequently, a total of 833 patches of SEAM Arid velocity model are obtained.

The first test only utilizes the classes as conditions for each patch. Figure \ref{fig:arid}c shows a sample of generated velocity models conditioned by FlatVelA class on all patches. We can observe a good continuity of the layers and also a general increase in velocity with depth. On the downside, the resolution of the generated velocity model is degraded and some artifacts are observed due to the features that are not within the learned distribution.

In the second test, we utilize both classes and a well as the conditions. We assume a well located at X = 416 in the velocity model. To incorporate the well velocities, we split the model at the well location (X = 416), flip the left part, predict both the left and right parts independently, and finally, concatenate them. These steps were conducted because of the prediction mechanism of our method that follows a particular direction; thus, to ensure the well information is utilized and propagated properly, we start predicting from the well location and then away in both directions. Figure \ref{fig:arid}d shows an example of the result of the second test. We can observe that the generated velocity model looks closer to the ground truth (Figure \ref{fig:arid}a) in terms of the layering.

Lastly, for the third test, we use classes and well conditions as in the second test. However, this time, we use a different class for patches located in the middle to show that we could still locally control the geological features in the generated model. Figure \ref{fig:arid}e shows an example of the generated velocity model with such a setting, where the middle layers are conditioned with the FlatFaultB class and the well is located at X = 0. We again observe the satisfactory alignment of the layering feature of the generated model with that of the ground truth (Figure \ref{fig:arid}a), with an additional faulting feature in the middle layers (Z = 100 -- 200).

\section{Discussions}
\label{sec:discussions}

The main purpose of this paper is to test the plausibility of using a top-down GPT implementation to build velocity models that can use the more constrained shallow part as prior to predicting the deep. For this proof of concept paper, we use small velocity models and simple examples. The VQ-VAE, as a compression (latent representation) tool is key to scaling up this approach to more practical velocity models and even 3D. However, we will now share some of the challenges of the approach. We observed some drawbacks based on the results in Figures \ref{fig:unconditional}--\ref{fig:local-pred}. A notable downside is the imperfect realization of the velocity models. As the GPT only handles the storage of the distribution of the velocity models, any issue related to resolution is attributed to the encoding/decoding process by the VQ-VAE. The VQ-VAE is optimized using a hybrid loss function (Equation 1) that penalizes the reconstruction and the size of the latent space. Therefore, there is a trade-off between the two depending on the weighting value (i.e., $\beta$), which is not trivial to choose. When $\beta$ is high (i.e., more weight on the commitment loss), the reconstruction quality may deteriorate. Conversely, when $\beta$ is low (i.e., more weight on the reconstruction loss), the latent size may grow arbitrarily. These phenomena could lead to the infamous problem of quantizing networks of \textit{codebook collapse}, in which the embedding is not optimally utilized \citep{takida2022sq}. Different optimization strategies were promoted to address this problem (e.g., \citealp{mentzer2023finite, takida2022sq}), which could be utilized in the future version of VelocityGPT.

Moreover, the simple architecture of the current VQ-VAE model, which is only a stack of convolutional layers, contributes to the reconstruction inaccuracy of the velocity models. This can be alleviated using a more advanced quantizing network (e.g., \citealp{yu2021vector}) with better reconstruction capability. Unlike in the natural images use case, preservation of edges (i.e., layer interfaces) is essential in the seismic domain, which provides accurate information on the depth of geological objects. Experimentation with techniques that could preserve the details of the interfaces in the velocity and post-stack image, like those proposed in \cite{lin2023catch}, is also a potential direction for an enhanced VelocityGPT.

Another issue is how to control the contribution of each prior/condition, which would allow more flexibility in controlling the generation of the velocity models. This feature is useful when, for example, VelocityGPT is utilized as a prior for an iterative inversion process. In an iterative inversion scheme (like FWI), early iterations commonly form the overall structure of a velocity model, in which to condition it dominantly with a class is appropriate. For later updates, finer-scale structures and layering details are recovered, in which the well condition could play a bigger role. Adjustable weights for each condition are also useful if one has a different level of confidence between the conditions.

\section{Conclusions}
\label{sec:conclusions}

We developed a framework called VelocityGPT to autoregressively generate velocity models from shallow subsurface to deep, by using the shallow part for guidance. A Vector-Quantized Variational Auto Encoder is used to allow the VelocityGPT to work on a reduced dimension representation of the velocity models. We trained VelocityGPT to accept conditions in the generation that includes geology class, well information and available images. Thus, we showed how different prior information of the velocity models, including the geological features (as classes), well-log velocities, and geological structural information (as post-stack images), can be incorporated during training to allow for conditional sampling in the inference stage. We demonstrated the network's capability to generate velocity models that agree with the injected priors. We also shared a mechanism to upscale the VelocityGPT to more realistic size models.

\section*{Acknowledgment}
This publication is based on work supported by the King Abdullah University of Science and Technology (KAUST). The authors thank the DeepWave sponsors for their support.

\bibliographystyle{plainnat}  
\bibliography{references}

\end{document}